\documentclass[peerreview,transmag,onecolumn]{IEEEtran}

\usepackage{amsmath}
\usepackage{amssymb}
\usepackage{bm}
\usepackage{multirow}
\usepackage{tabularx}
\usepackage{graphicx}
\usepackage{cite}
\usepackage{textcomp}

\begin{document}
\renewcommand*\listfigurename{Figure captions}

\title{Gaussian process regression for forest attribute estimation from airborne laser scanning data}
\author{Petri~Varvia,
Timo~L\"{a}hivaara,
Matti~Maltamo,
Petteri~Packalen,
Aku~Sepp\"{a}nen%
\thanks{This work was supported by the Finnish Cultural Foundation, North Savo Regional fund, the Academy of Finland (Project numbers 270174, 295341, 295489, and 303801, and Finnish Centre of Excellence of Inverse Modelling and Imaging 2018-2025), and the FORBIO project  (The Strategic Research Council, Grant No. 293380).
}%
\thanks{P. Varvia was with the Department of Applied Physics, University of Eastern Finland, FI-70211 Kuopio, Finland. He is now with the Laboratory of Mathematics, Tampere University of Technology, FI-33101 Tampere, Finland (e-mail: petri.varvia@gmail.com).}%
\thanks{T. L\"{a}hivaara and A. Sepp\"{a}nen are with the Department of Applied Physics, University of Eastern Finland, FI-70211 Kuopio, Finland.}%
\thanks{M. Maltamo and P. Packalen are with the School of Forest Sciences, University of Eastern Finland, FI-80101 Joensuu, Finland.}}

\maketitle

\begin{abstract}
While the analysis of airborne laser scanning (ALS) data often provides reliable estimates for certain forest stand attributes -- such as total volume or basal area -- there is still room for improvement, especially in estimating species-specific attributes. Moreover, while information on the estimate uncertainty would be useful in various economic and environmental analyses on forests, a computationally feasible framework for uncertainty quantifying in ALS is still missing. In this article, the species-specific stand attribute estimation and uncertainty quantification (UQ) is approached using Gaussian process regression (GPR), which is a nonlinear and nonparametric machine learning method. Multiple species-specific stand attributes are estimated simultaneously: tree height, stem diameter, stem number, basal area, and stem volume. The cross-validation results show that GPR yields on average an improvement of 4.6\% in estimate RMSE over a state-of-the-art k-nearest neighbors (kNN) implementation, negligible bias and well performing UQ (credible intervals), while being computationally fast. The performance advantage over kNN and the feasibility of credible intervals persists even when smaller training sets are used.

\end{abstract}

\tolerance 1000
\section*{Copyright notice}
P. Varvia, T. L\"{a}hivaara, M. Maltamo, P. Packalen and A. Sepp\"{a}nen, "Gaussian Process Regression for Forest Attribute Estimation From Airborne Laser Scanning Data," in IEEE Transactions on Geoscience and Remote Sensing. doi: 10.1109/TGRS.2018.2883495 \\

\textcopyright 2018 IEEE.  Personal use of this material is permitted.  Permission from IEEE must be obtained for all other uses, in any current or future media, including reprinting/republishing this material for advertising or promotional purposes, creating new collective works, for resale or redistribution to servers or lists, or reuse of any copyrighted component of this work in other works.

\section{Introduction}
Forest inventories based on airborne laser scanning (ALS) are becoming increasingly popular. Therefore, it is more and more important to have well performing methods for the estimation/prediction of stand attributes, such as basal area and tree height. Coupled with the prediction procedures, efficient methods for the quantification of prediction uncertainty are also urgently needed for forestry planning and assessment purposes \cite{kangas2018}. 

Operational forest inventories employing ALS data are most often implemented with the area based approach (ABA) \cite{naesset2002}. In ABA, metrics used as predictor variables are calculated from the ALS returns within a plot or grid cell. Using training plots with field-measured stand attributes, a model is formulated between the stand attributes and ABA metrics. This statistical model is then used to predict the stand attributes for each grid cell \cite{reutebuch2005,maltamobook} and the predictions are finally aggregated to the desired area, e.g. to a stand. Although tree species is among the most important attributes of forest inventory, the ALS research does not particularly reflect this. One reason for this is that in many biomes the number of tree species is so high that it is practically impossible to separate them by remote sensing. In the Nordic countries, however, the majority of the growing stock comes from three economically valuable tree species. The species-specific prediction is approached two ways in Nordic countries: in Norway, stands are stratified according to tree species by visual interpretation of aerial images before the actual ALS inventory \cite{naesset2004}, whereas in Finland, stand attributes are predicted by tree species using a combined set of metrics from ALS data and aerial images \cite{packalen2007}. In both approaches, aerial images are used to improve the discrimination of tree species.

Uncertainty estimation is a key component in strategic inventories that cover large areas \cite{mandallaz2007}. ALS can be used in that context too. For example, ALS metrics can be used as auxiliary variables in model-based (e.g. \cite{staahl2010}) or model assisted (e.g. \cite{gregoire2010}) estimation of some forest parameter. Typically, sample mean and sample variance are estimated to the area of interest (e.g. 1000000 ha) using a certain number (e.g. 500) of sample plots and auxiliary variables covering all population elements. In the stand level forest management inventories, the situation is different: the point estimate and its confidence intervals are needed for each stand and there may not be any sample plots in most stands. Today, most ALS inventories can be considered as stand level management inventories.

Commonly in ABA, when using prediction methods such as linear regression or \mbox{kNN}, only point estimates without accompanying uncertainty metrics are computed. Plot or cell level prediction uncertainty has garnered some research interest in recent years and several methods of predicting plot/cell level variance have been proposed \cite{junttila2008a,finley2013,magnussen2016}. Recently, a Bayesian inference approach to quantify uncertainty within the framework of the ABA was proposed by Varvia \emph{et al.} \cite{varvia}. The main shortcoming in the method proposed in \cite{varvia} is that it is computationally costly: wall-to-wall uncertainty quantification of a large forest area would require considerable computer resources.

Gaussian process regression (GPR) \cite{rasmussenbook} is a machine learning method that provides an attractive alternative; compared with the more widely used machine learning methods, such as artificial neural networks \cite{niska2010neural,alsdeeplearning}, GPR also produces an uncertainty estimate for the prediction. Univariate GPR was tested for estimation of several total stand attributes by Zhao \emph{et al.} \cite{alsgpr}, where it was found to significantly outperform (log)linear regression.

In this paper, we propose a multivariate GPR for simultaneous estimation of species-specific stand attributes within ABA. The estimation accuracy of GPR is compared with kNN and the uncertainty quantification performance with the Bayesian inference method of \cite{varvia}. Furthermore, the effect of training set size on its performance is evaluated.

\section{Materials}
The same test data as in \cite{varvia} is used in this study. In this section, the data set is briefly summarized, for detailed description, see e.g. \cite{packalen2009,Packalen2012}. The test area is a managed boreal forest located in Juuka, Finland. The area is dominated by Scots pine (\emph{Pinus sylvestris} L.) and Norway spruce (\emph{Picea abies} (L.) Karst.), with a minority of deciduous trees, mostly downy birch (\emph{Betula pubescens} Ehrh.) and silver birch (\emph{Betula pendula} Roth.). The deciduous trees are considered as a single group. 

The field measurements were done during the summers of 2005 and 2006. Total of 493 circular sample plots of radius 9 m are used in this study. The diameter at breast height (DBH), tree and storey class, and tree species were recorded for each tree with DBH larger than 5 cm and the height of one sample tree of each species in each storey class was measured. The heights of other trees on the plot were predicted using a fitted N\"aslund's height model \cite{naslund}. The species-specific stand attributes were then calculated using the measured DBH and the predicted heights. The stand attributes considered in this study are tree height ($H_{\mathrm{gm}}$), diameter at breast height ($D_{\mathrm{gm}}$), stem number ($N$), basal area ($\mathit{BA}$), and stem volume ($V$). 

The ALS data and aerial images were captured in 13 July 2005 and 1 September 2005, respectively. The ALS data has a nominal sampling density of 0.6 returns per square meter, with a footprint of about 60 cm at ground level. The orthorectified aerial images contain four channels (red, green, blue, and near infrared). A total of $n_x=77$ metrics were computed from the ALS point cloud and aerial images and used in ABA. The metrics include canopy height percentiles, the corresponding proportional canopy densities, the mean and standard deviation of the ALS height distribution, the fraction of above ground returns (i.e. returns with $z>2$ m), and metrics computed from the LiDAR intensity. From the aerial images, the mean values of each channel were used along with two spectral vegetation indices \cite{packalen2009}.

\section{Methods}
\label{sec:methods}
Let us denote a vector consisting of the stand attributes by $\mathbf{y}\in\mathbb{R}^{15}$; the vector $\mathbf{y}$ contains the species-specific (pine, spruce, deciduous) $H_{\mathrm{gm}}$, $D_{\mathrm{gm}}$, $N$, $\mathit{BA}$, and $V$, resulting in a total of $n_y=15$ variables. The vector of predictors (ALS and aerial image metrics) is denoted by $\mathbf{x}\in\mathbb{R}^{n_x}$. 

The general objective is to learn a nonlinear regression model
\begin{equation}
\label{thefunc}
\mathbf{y}=f(\mathbf{x})+\mathbf{e},
\end{equation}
where $\mathbf{e}$ is an error term, from a set of $n_t$ training data $(\mathbf{Y}_t,\mathbf{X}_t)$.

Let $\mathbf{Y}$ be a finite collection of points $\mathbf{y}^{(i)}=f(\mathbf{x}^{(i)})+\mathbf{e}^{(i)}$ concatenated in a long vector. In Gaussian process regression \cite{rasmussenbook}, the joint probability distribution of these points $\mathbf{Y}$ is modeled as a multivariate normal distribution, with mean $\boldsymbol{\mu}_{\mathbf{y}}$  and covariance $\boldsymbol{\Gamma}_{\mathbf{y}}$ written as functions of $\mathbf{x}$:
\begin{align}
&\boldsymbol{\mu}_{\mathbf{y}} = \mathbf{m}(\mathbf{x}) = \mathbb{E}\{f(\mathbf{x})\},\\
&\boldsymbol{\Gamma}_{\mathbf{y}}=\mathbf{K}(\mathbf{x},\mathbf{x}^{\prime}) = \mathbb{E}\{(f(\mathbf{x})-\mathbf{m}(\mathbf{x}))(f(\mathbf{x}^{\prime})-\mathbf{m}(\mathbf{x}^{\prime}))^T\},
\end{align}
where $(\:\cdot\:)^T$ is the matrix transpose. Let now $\mathbf{Y} = \begin{bmatrix} \mathbf{Y}_t & \mathbf{y}_*\end{bmatrix}^T$, that is, $\mathbf{Y}$ a vector consisting of the training data $\mathbf{Y}_t$, and a new point $\mathbf{y}_*$ which we want to estimate, using the corresponding measurement $\mathbf{x}_*$. For simplification, we set $\mathbf{m}(\mathbf{x})=0$. The mean term mostly affects the behavior when extrapolating far away from the space covered by the training data. The joint distribution of $\mathbf{Y}$ is then
\begin{equation}
\label{joint}
\begin{bmatrix} \mathbf{Y}_t \\ \mathbf{y}_* \end{bmatrix} \sim\mathcal{N}\left(0,
\begin{bmatrix} \mathbf{K}(\mathbf{X}_t,\mathbf{X}_t)+\mathbf{E} & \mathbf{K}(\mathbf{x}_*,\mathbf{X}_t)^T \\ \mathbf{K}(\mathbf{x}_*,\mathbf{X}_t) & \mathbf{K}(\mathbf{x}_*,\mathbf{x}_*)+\mathbf{E}_* \end{bmatrix}\right),
\end{equation}
where $\mathbf{E}$ and $\mathbf{E}_*$ describe the covariance of the error $\mathbf{e}$, i.e. uncertainty of $\mathbf{y}$. In this work, we use $\mathbf{E}*=0.1\mathbf{D}$, where $\mathbf{D}$ is a diagonal matrix that contains the sample variances of the training data $\mathbf{Y}_t$ on the main diagonal. The error matrix $\mathbf{E}= 0.1\mathbf{D}\otimes \mathbf{I}$, where $\otimes$ is the Kronecker product and $\mathbf{I}\in\mathbb{R}^{n_t\times n_t}$ is an identity matrix. For brevity, following shorthand notations are introduced: 
\begin{align}
&\mathbf{K} = \mathbf{K}(\mathbf{X}_t,\mathbf{X}_t)\in\mathbb{R}^{n_yn_t\times n_yn_t} \\
&\mathbf{K}_* =  \mathbf{K}(\mathbf{x}_*,\mathbf{X}_t)\in\mathbb{R}^{n_y\times n_yn_t}.
\end{align}

In GPR, the kernel matrices $\mathbf{K}$ and $\mathbf{K}_* $ are constructed based on a covariance function. In this study we use stationary Mat\'{e}rn covariance function with $\nu=3/2$, fixed length scale $l=10$, and $\sigma=1$:
\begin{equation}
\label{matern}
k(\mathbf{x},\mathbf{x^\prime}) = \left(1+\frac{\sqrt{3}d(\mathbf{x},\mathbf{x^\prime})}{10}\right)\exp\left(-\frac{\sqrt{3}d(\mathbf{x},\mathbf{x^\prime})}{10}\right),
\end{equation}
where the distance metric $d(\mathbf{x},\mathbf{x^\prime})$ is the Euclidean distance. The covariance function $k(\mathbf{x},\mathbf{x^\prime})$ describes the covariance between the vectors $\mathbf{x}$ and $\mathbf{x^\prime}$ based on the distance between the vectors. The covariance function is the core component of GPR that specifies properties such as smoothness of the regressor.

The covariance function \eqref{matern} is used to construct univariate kernel matrices
\begin{align}
&K(i,j) = k(\mathbf{x}_t^{(i)},\mathbf{x}_t^{(j)})\in\mathbb{R}^{n_t\times n_t} \\
&K_*(1,j) = k(\mathbf{x}^*,\mathbf{x}_t^{(j)})\in\mathbb{R}^{1\times n_t}.
\end{align}
To get from the univariate kernels to multivariate kernels used in \eqref{joint}, the so-called separable kernel \cite{bonilla2008} is used:
\begin{align}
&\mathbf{K} = \boldsymbol{\Gamma}_{\mathbf{y}}\otimes K \\
&\mathbf{K}_* = \boldsymbol{\Gamma}_{\mathbf{y}}\otimes K_*,
\end{align}
where $\boldsymbol{\Gamma}_{\mathbf{y}}\in\mathbb{R}^{n_y\times n_y}$ is a (prior) covariance for $\mathbf{y}$. In this work, $\boldsymbol{\Gamma}_{\mathbf{y}}$ is approximated by the sample covariance of $\mathbf{Y}_t$. It should be noted, that $\sigma=1$ is chosen in the kernel function \eqref{matern}, because the (prior) variances of $\mathbf{y}$ are added to the covariance kernel in this step through $\boldsymbol{\Gamma}_{\mathbf{y}}$.

From the joint density \eqref{joint}, the conditional density of $\mathbf{y}_*$ given the training data and the measurement $\mathbf{x}_*$ is
\begin{equation}
\mathbf{y}_*\:|\:\mathbf{Y}_t,\mathbf{X}_t,\mathbf{x}_*\sim\mathcal{N}(\mathbf{m}(\mathbf{y}_*),\boldsymbol{\Gamma}_{\mathbf{y}_*}),
\end{equation}
where
\begin{align}
\label{predmean}
&\boldsymbol{\mu}_{\mathbf{y}_*} = \mathbf{K}_*(\mathbf{K}+\mathbf{E})^{-1}\mathbf{y}_t \\
\label{predcov}
&\boldsymbol{\Gamma}_{\mathbf{y}_*}= \boldsymbol{\Gamma}_{\mathbf{y}} + \mathbf{E}_* - \mathbf{K}_*(\mathbf{K}+\mathbf{E})^{-1}\mathbf{K}_*^T.
\end{align}
The predictive mean $\boldsymbol{\mu}_{\mathbf{y}_*} $ is now the point estimate for the unknown vector of stand attributes $\mathbf{y}_*$ and $\boldsymbol{\Gamma}_{\mathbf{y}_*}$ provides the estimate covariance. As can be seen from the equation \eqref{predmean}, the final prediction is a linear combination of the training data values $\mathbf{y}_t$. This mathematical connection to linear models is expected, because general linear model can be written as a special case of GPR \cite{rasmussenbook}.

\subsection{Correcting for negative predictions}
Unlike kNN and certain other machine learning methods, GPR extrapolates outside the training data. As an unwanted side effect of this extrapolation behavior, GPR can produce unrealistic negative predictions for the stand attributes. Several statistically rigorous methods for constraining the GPR predictions have been proposed \cite{DaVeiga2012,jidling2017}, but these methods increase the computational cost significantly and are nontrivial to implement. Here we adopt a simpler correction.

For the point prediction, we compute a maximum a posteriori estimate by solving
\begin{equation}
\hat{\mathbf{y}}_*=\mathrm{arg}\:\underset{\hat{\mathbf{y}}}{\mathrm{min}}\left\{(\hat{\mathbf{y}}-\boldsymbol{\mu}_{\mathbf{y}_*})^T\boldsymbol{\Gamma}_{\mathbf{y}_*}^{-1}(\hat{\mathbf{y}}-\boldsymbol{\mu}_{\mathbf{y}_*})\right\},\;\hat{\mathbf{y}}\geq 0.
\end{equation}
The prediction $\hat{\mathbf{y}}_*$ is the mode of the truncated GPR predictive density. If the original GPR predictive mean is non-negative, $\hat{\mathbf{y}}_*$ is simply $\boldsymbol{\mu}_{\mathbf{y}_*}$.

Due to the complicated structure of the marginal densities of a truncated multivariate Gaussian distribution \cite{Horrace2005}, correcting the predictive intervals exactly is not computationally practical. Instead, the univariate Gaussian marginals of the GPR predictive density are truncated at zero. If the original 95\% predictive interval for a stand attribute is $[a,b]$ and $a<0$, set $\hat{a}=0$ and calculate the new corrected upper bound $\hat{b}$ using the cumulative distribution of univariate truncated Gaussian by solving
\begin{equation}
\Phi(\hat{b},\mu_{y_{*}},\sigma_{y_{*}}) = 0.95+0.05\Phi(0,\mu_{y_{*}},\sigma_{y_{*}}),
\end{equation}
where $\Phi(\:\cdot\:,\mu_{y_{*}},\sigma_{y_{*}})$ is the cumulative distribution function of the univariate Gaussian distribution with the mean and standard deviation from the GPR predictive distribution. If $a\geq0$, the interval $[a,b]$ does not change. The corrected interval $[\hat{a},\hat{b}]$ is not a proper predictive interval of the truncated predictive distribution, unless $\boldsymbol{\Gamma}_{\mathbf{y}_*}$ is strictly diagonal.

\subsection{Reference methods}
The GPR point estimates are compared with a state-of-the-art kNN algorithm. We select ten predictors from the (transformed) data using a simulated annealing -based optimization approach of \cite{Packalen2012} and use the most similar neighbor (MSN) method for selecting the neighbors. The number of neighbors is chosen to be $k=5$, as in \cite{packalen2009,Packalen2012}. The predictor selection is done using the whole data set and leave-one-out cross-validation.

The prediction credible intervals provided by GPR are compared with the Bayesian inference approach \cite{varvia}. In the Bayesian approach the posterior predictive density:
\begin{equation}
\label{alsextposterior}
\pi(\mathbf{y}_*|\mathbf{x}) \propto \begin{cases} \mathcal{N}(\mathbf{x}|\hat{\mathbf{A}}\boldsymbol{\phi}(\mathbf{y}_*)+\hat{\boldsymbol{\mu}}_{\mathbf{e}|\mathbf{y}},\hat{\boldsymbol{\Gamma}}_{\mathbf{e}|\mathbf{y}}) &\\
\qquad\qquad\quad\;\cdot\:\mathcal{N}(\mathbf{y}_*|\hat{\boldsymbol{\mu}}_{\boldsymbol{\theta}},\hat{\boldsymbol{\Gamma}}_{\mathbf{y}}), & \mathbf{y}_*\geq 0 \\
0, & \mathbf{y}_*< 0, \end{cases}
\end{equation}
is constructed based on the training data and the new measurement. The model matrix $\hat{\mathbf{A}}$, conditional (residual) error statistics $\hat{\boldsymbol{\mu}}_{\mathbf{e}|\mathbf{y}}$ and $\hat{\boldsymbol{\Gamma}}_{\mathbf{e}|\mathbf{y}}$, and the prior statistics $\hat{\boldsymbol{\mu}}_{\boldsymbol{\theta}}$ and $\hat{\boldsymbol{\Gamma}}_{\mathbf{y}}$ are learned from the training data. The density \eqref{alsextposterior} is then sampled using a Markov chain Monte Carlo method. The point estimate and 95\% credible intervals are then calculated from the samples.

\begin{figure*}[htb]
	\centering
	\includegraphics[width=\textwidth]{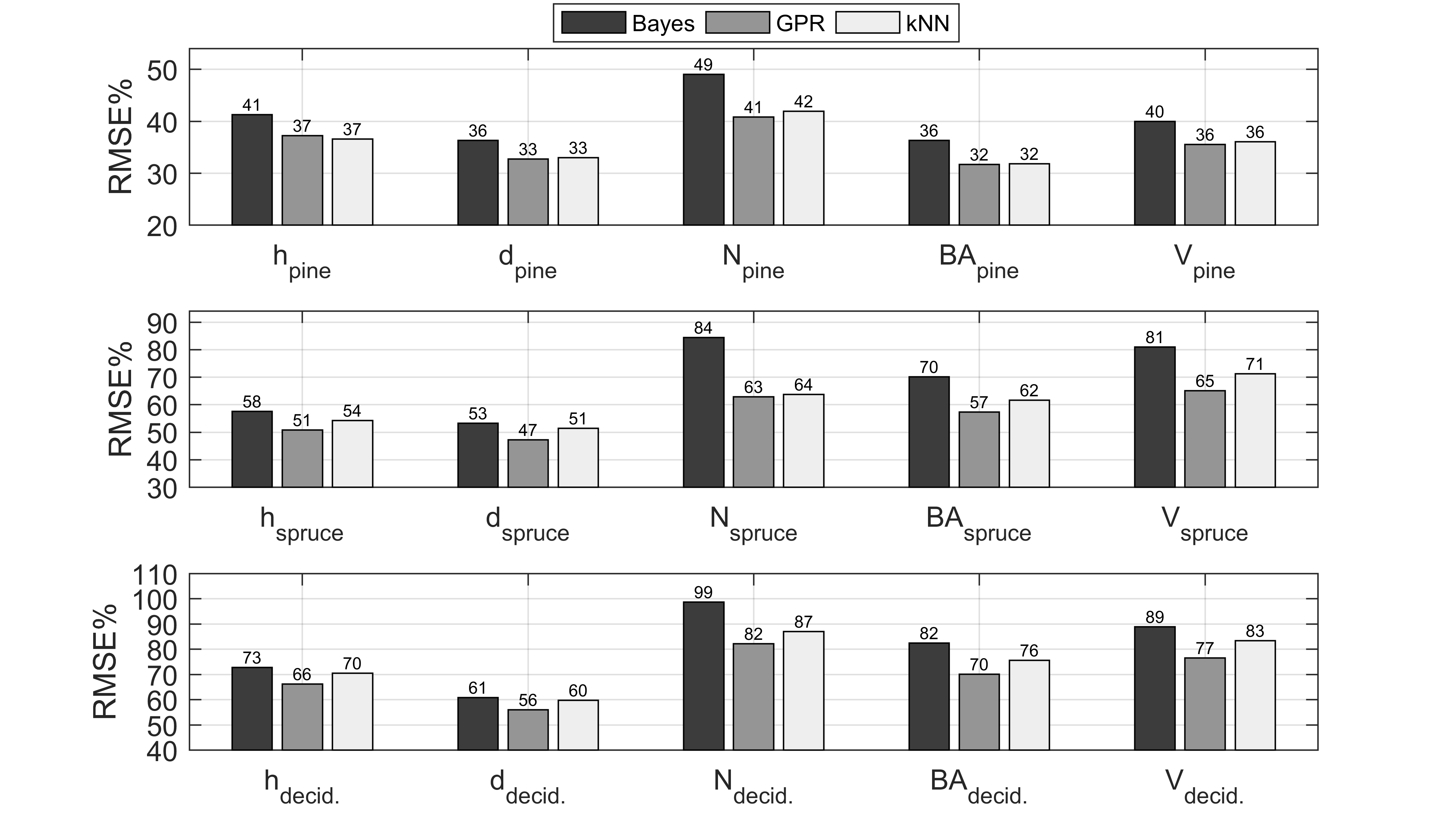}
	\caption{Relative RMSE of Bayesian linear, GPR, and kNN estimates for pine (top), spruce (middle), and deciduous (bottom). Numerical values are shown on top of the bars.}
	\label{fig:rmse}
\end{figure*}

\subsection{Performance assessment}
The proposed GPR method is first evaluated using leave-one-out cross-validation (i.e. $n_t=492$). From the results, relative root mean square error (RMSE\%), relative bias (bias\%), and credible interval coverage (CI\%) are calculated. Credible interval coverage is the percentage of the test plots where the field measured value of a stand attribute lies inside the computed 95\% prediction interval; CI\% thus has the ideal value of 95\%.

In addition to conducting a leave-one-out cross-validation, the effect of the number of training plots is evaluated. Training set sizes from $n_t=20$ to $n_t=400$ are tested with a stepping of 20. The cross-validation is performed by first randomly sampling $n_t$ plots to be used as a training set and then randomly selecting a single test plot from the remaining $493-n_t$ plots. This procedure is repeated 2000 times for each $n_t$ value. This way the number samples for each tested $n_t$ stays constant. The effect of training set size is only evaluated for GPR and kNN, due to the high computational cost of the reference Bayesian inference approach.

\section{Results and discussion}
\label{sec.results}
\subsection{Species-specific attributes}
The RMSE\% comparison between the GPR predictions, kNN and Bayesian inference is shown in the Figure \ref{fig:rmse} for all estimated stand attributes. The numerical RMSE\% value is shown above each bar. For pine, which is the dominant species in the study area, the GPR and kNN estimates have fairly equivalent performance: GPR is slightly better for all the stand attributes except height and basal area. The Bayesian linear inference estimates are notably worse. In the minority species (spruce and deciduous), the GPR estimates have consistently better RMSE\% than kNN or Bayesian linear. On average, the relative improvement over kNN is 6.5\% for the minority species and 4.6\% for all species.

\begin{figure*}[htb]
	\centering
	\includegraphics[width=\textwidth]{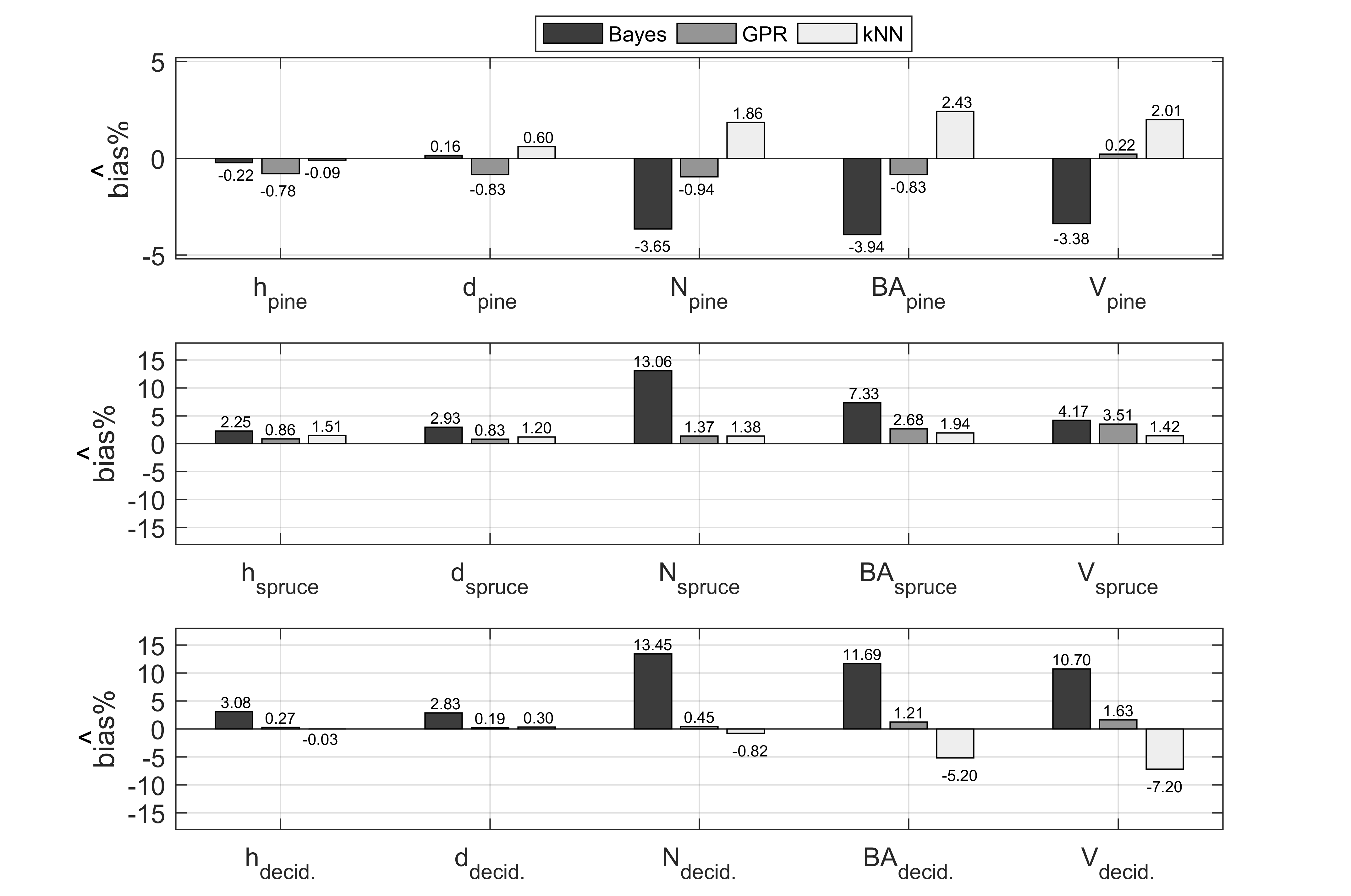}
	\caption{Relative bias of Bayesian linear, GPR, and kNN estimates for pine (top), spruce (middle), and deciduous (bottom). Numerical values are shown on top of the bars.}
	\label{fig:bias}
\end{figure*}

Figure \ref{fig:bias} shows a similar comparison of relative bias between the evaluated methods for all the estimated stand attributes.  The numerical bias\% value is printed for each bar. GPR estimates show smaller than 2\% absolute bias for all the stand attributes, except the spruce basal area and volume. kNN shows small bias in the spruce attributes, but has a large bias in deciduous basal area and volume. The Bayesian linear results show notable bias in $N$, $\mathit{BA}$, and $V$.

\begin{figure*}[htb]
	\centering
	\includegraphics[width=140mm]{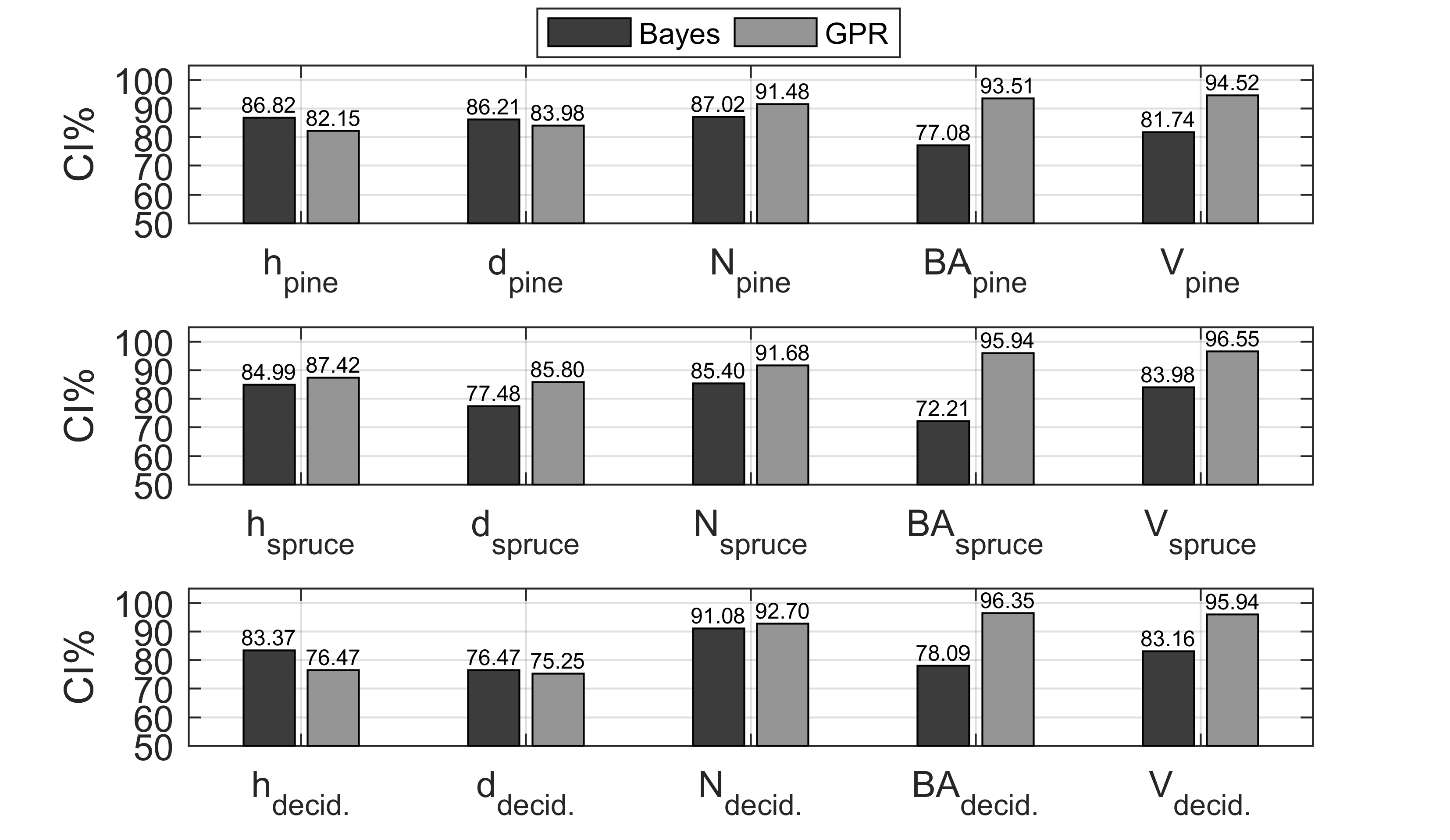}
	\caption{CI\% of Bayesian linear and GPR estimates for pine (top), spruce (middle), and deciduous (bottom). Numerical values are shown on top of the bars.}
	\label{fig:ci}
\end{figure*}

The CI coverages of GPR and the reference Bayesian inference method are compared in Figure \ref{fig:ci}. The numerical CI\% value is shown above each bar; the ideal value is here 95\%.
The CI\% for the Bayesian linear estimates fall short of the 95\% target, that is, the prediction intervals that are too narrow. The GPR prediction intervals perform well on basal area and stem volume, with good coverage also on stem number. The GPR CI\% for these stand attributes is consistently better than the Bayesian linear. The GPR prediction intervals for height and diameter are overconfident, especially in the deciduous variables, and the performance is roughly similar to the Bayesian linear estimates.

\subsection{Total attributes}
Point estimates and credible intervals for the total stem number, basal area, and stem volume were calculated from the species-specific results. The point estimates were computed by summing up the corresponding species-specific estimates. The GPR prediction interval for the total attributes is acquired from the prediction covariance $\boldsymbol{\Gamma}_{\mathbf{y}_*}$, because summation is a linear transformation.

\begin{figure*}[htb]
	\centering
	\includegraphics[width=140mm]{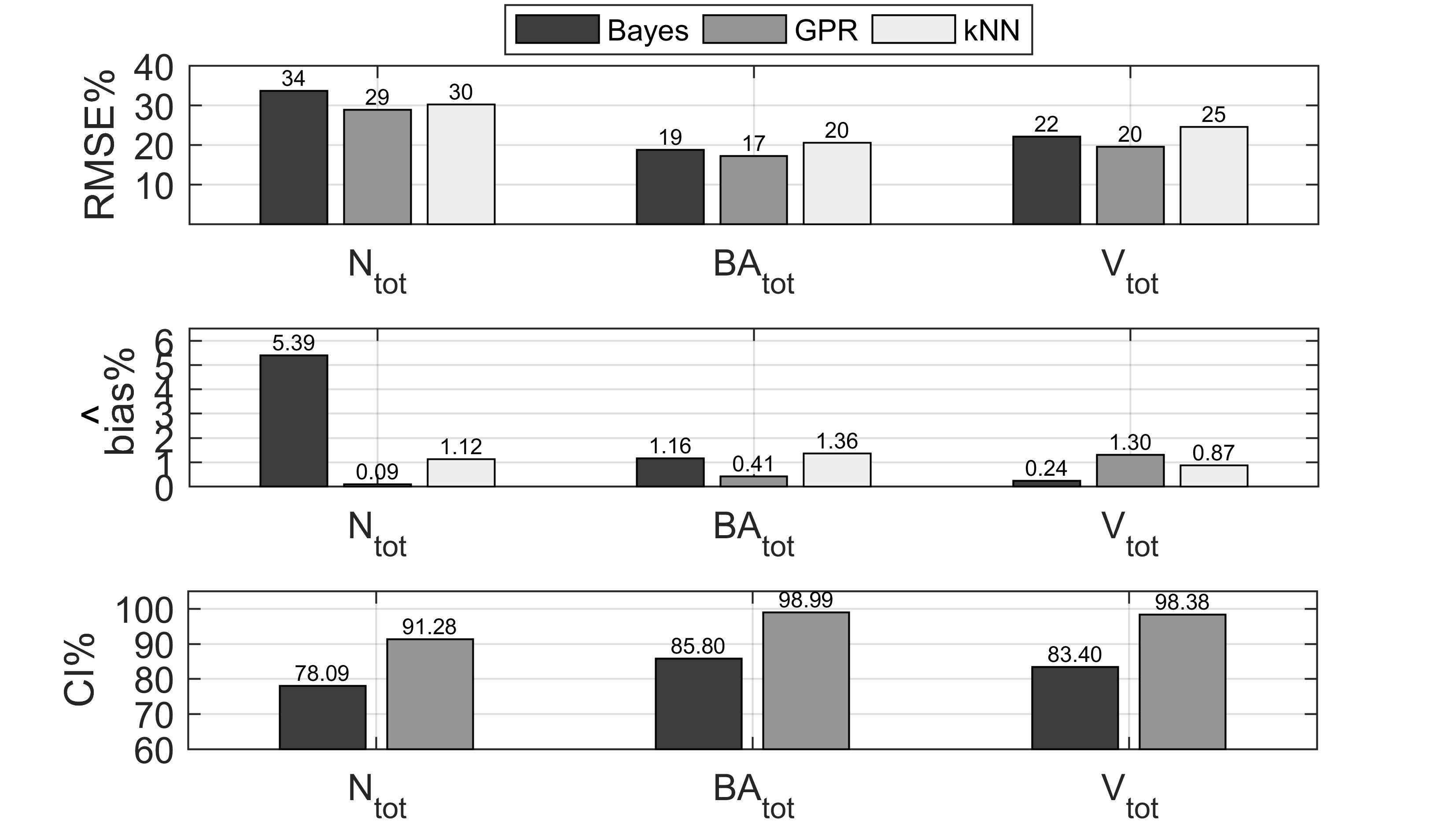}
	\caption{RMSE\% (top), bias\% (middle), and CI\% (bottom) of the tested methods for the total stand attributes. Numerical values are shown on top of the bars.}
	\label{fig:totvars}
\end{figure*}

The results for the total attributes are shown in Figure \ref{fig:totvars}. In RMSE\%, GPR estimates show the best performance. Bayesian linear estimates have lower RMSE\% than kNN in the total basal area and volume, while kNN is better in the stem number. In the relative bias, GPR has fairly low bias and performs worst in the total stem volume. kNN has consistent slight bias, while the Bayesian linear estimates show large bias in the stem number. In credible interval coverage, GPR produces too wide intervals for basal area and stem volume (CI\% between 98-99\%). The Bayesian linear intervals are, on the other hand, with CI\% around roughly 80\%.

\subsection{Effect of training set size}
\begin{figure}[htb]
	\centering
	\includegraphics[width=70mm]{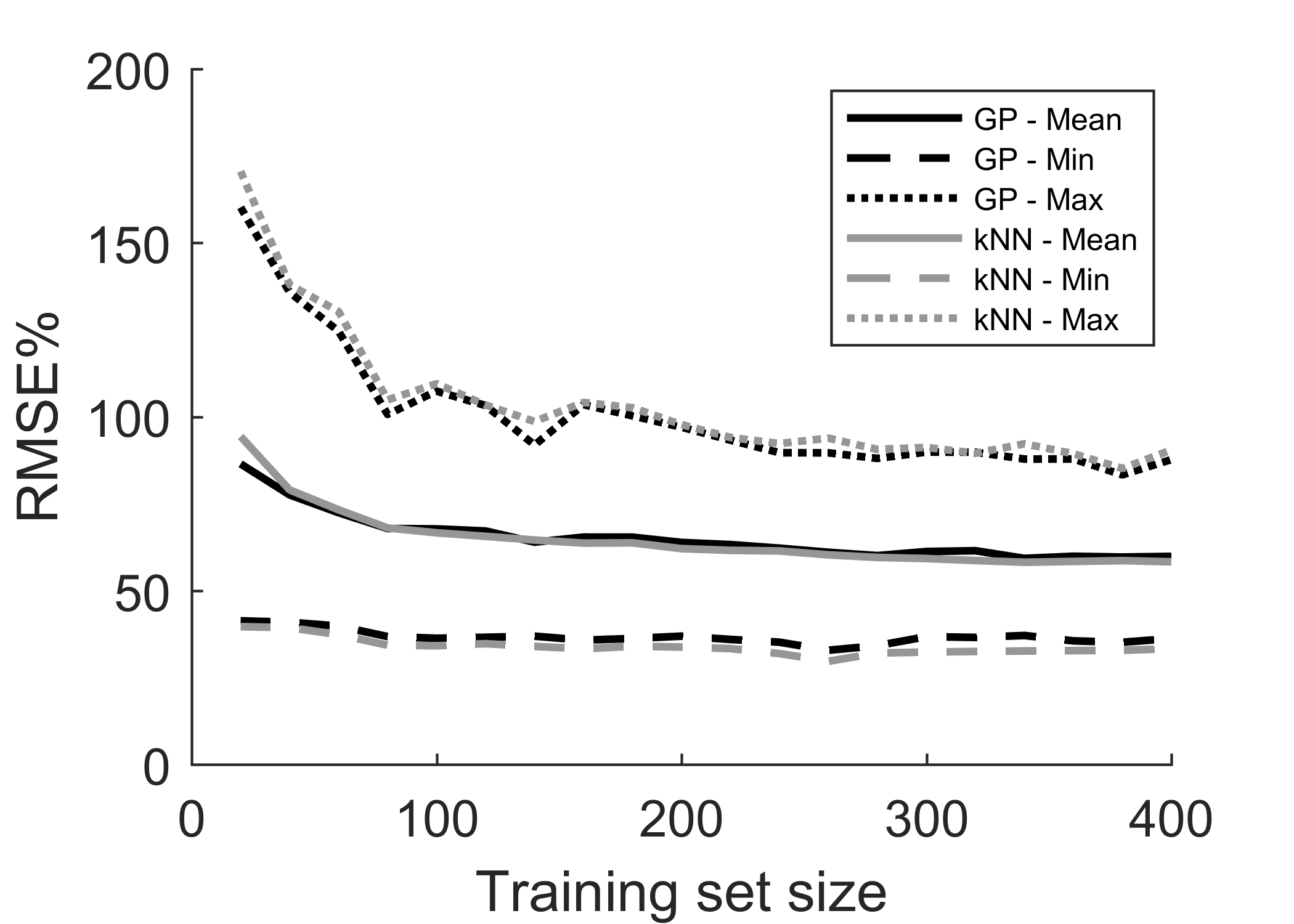}
	\caption{Lowest, average, and the highest relative RMSE as a function of training set size for GPR and kNN estimates.}
	\label{fig:rmsetrain}
\end{figure}

RMSE\% versus training set size is shown in Figure \ref{fig:rmsetrain}. The dashed minimum line corresponds to the species-specific stand attribute with the lowest RMSE\%, maximum to the stand attribute with the highest RMSE\%, and mean is the average over the stand attributes. As expected, the RMSE\% increases for both methods when the training set size decreases. GPR keeps the slight performance advantage over kNN even when using smaller training sets. The improvement in performance for training sets larger than c. 200 plots is fairly small.

\begin{figure}[htb]
	\centering
	\includegraphics[width=70mm]{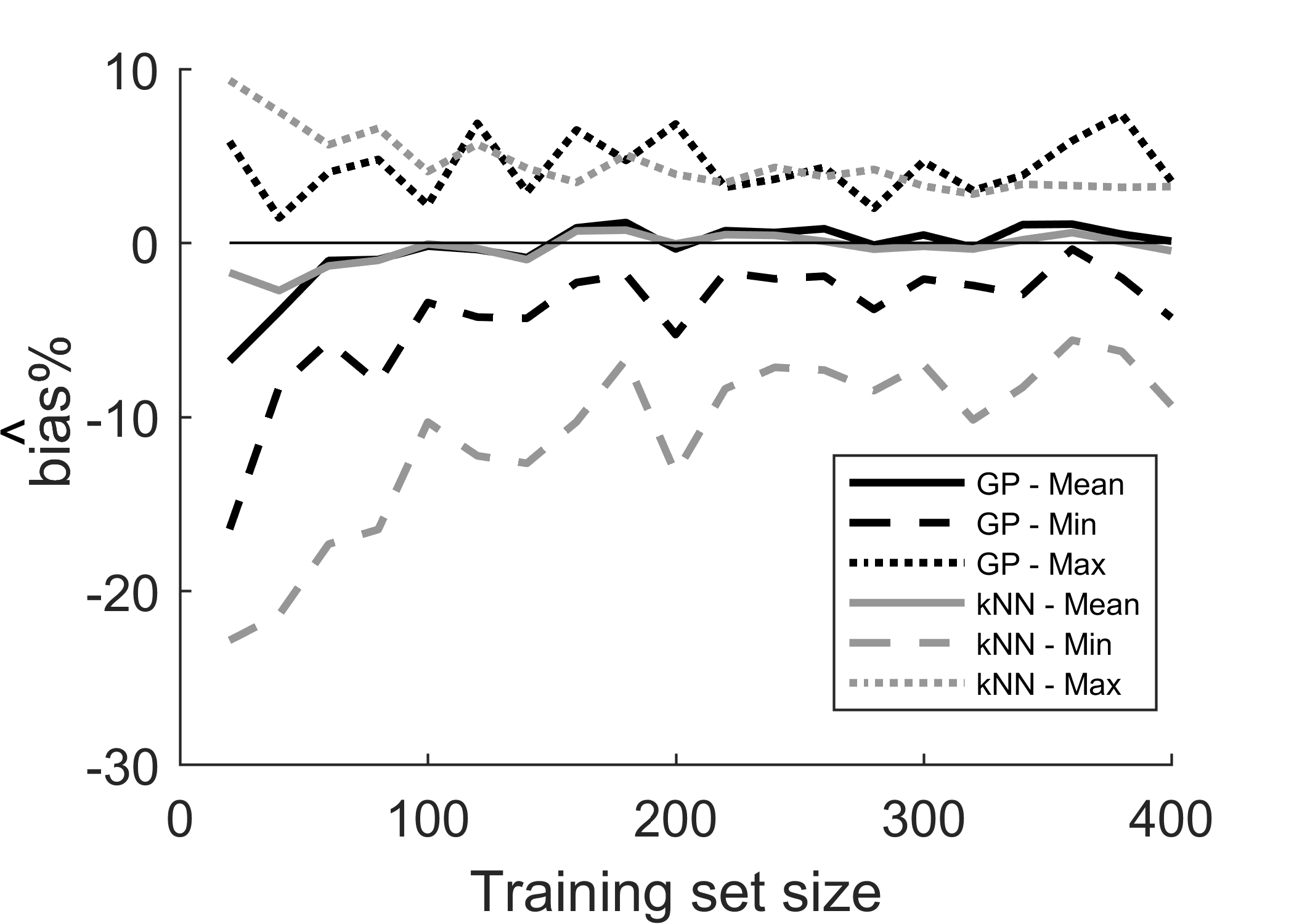}
	\caption{Lowest, average, and the highest relative bias as a function of training set size for GPR and kNN estimates.}
	\label{fig:biastrain}
\end{figure}

Figure \ref{fig:biastrain} shows the relative bias as a function of the training set size. When the training set size decreases, the estimated bias increases in both positive and negative directions, but with a general negative tendency. Smaller training sets are less likely to cover the full range of variation of the stand attributes in the population, which results in underestimation of large values: this would explain the observed tendency in bias. The largest negative biases produced by kNN are consistently larger than in GPR.

\begin{figure}[htb]
	\centering
	\includegraphics[width=70mm]{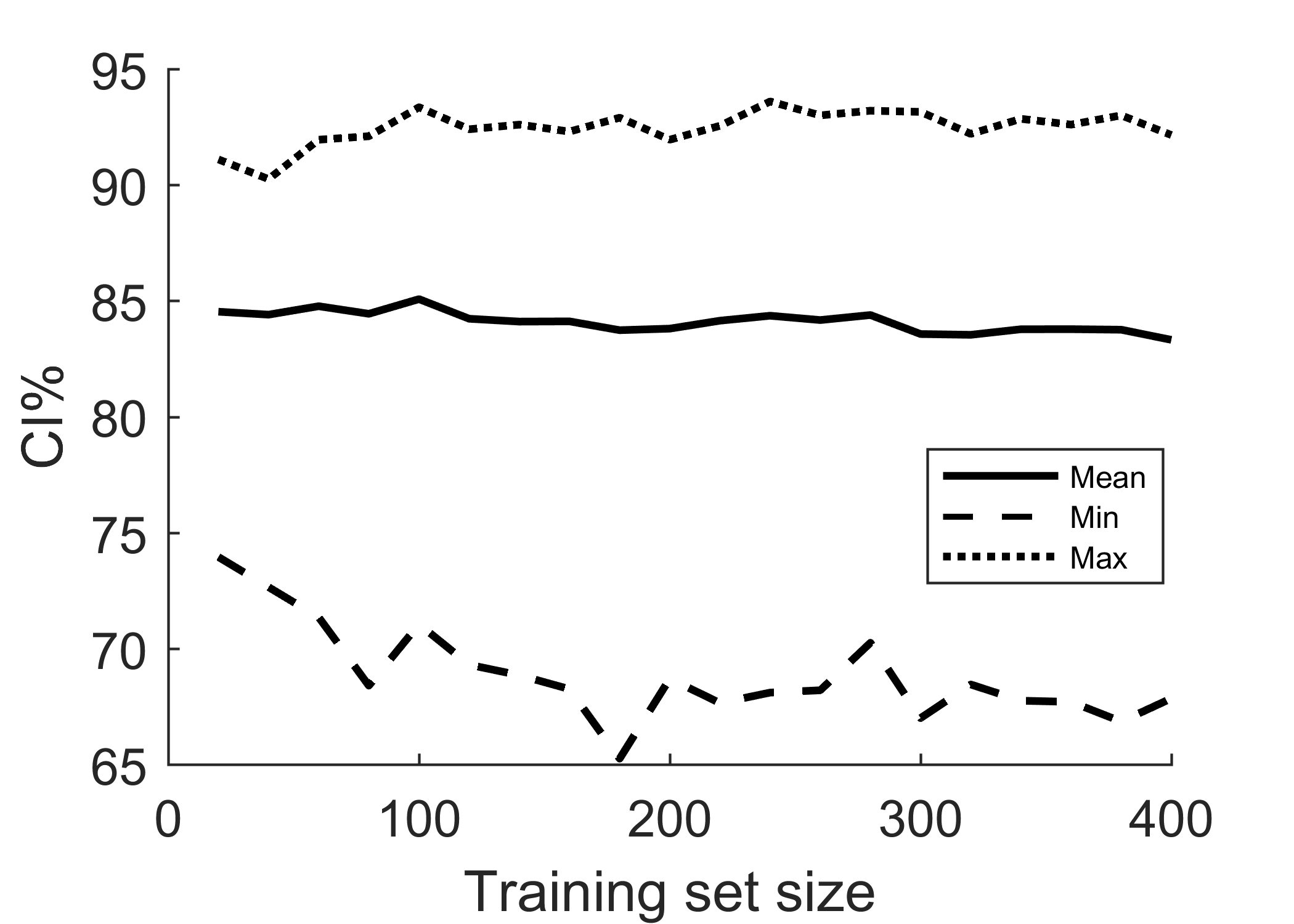}
	\caption{Lowest, average, and the highest CI\% as a function of training set size for GPR estimates.}
	\label{fig:citrain}
\end{figure}

Figure \ref{fig:citrain} shows the CI\% of the GPR estimates versus the training set size. The average CI\% increases slightly as training set size is decreased, the lowest CI\% increases considerably, while the highest CI\% drops somewhat. The generally too narrow credible intervals signify overconfidence in the predictions, which implies that either the GP model is not optimal in its current formulation, or the stand attributes have not sufficiently explained the variation in the predictors. The latter explanation might cause that there are usually contradicting training data (i.e. training points that are close in the stand attribute space, but distant in the predictor space) in large training sets, which might partly explain the slight improvement of CI\% when training set size decreases. With the Bayesian inference approach, on the contrary, a substantial drop in CI\% in smaller training set sizes would be expected based on the results in \cite{varvia}.

\subsection{Discussion}
\label{sec:disc}
Conceptually, GPR is a non-parametric machine learning method that has similarities with kNN. Thus, many approaches proposed for improvement of kNN estimates within ABA could be also utilized to further improve GPR estimates. GPR seems to be insensitive to multicollinearity and quite large numbers of predictors can been used simultaneously \cite{alsgpr}. In this paper, fairly traditional ABA metrics were used, adding additional predictors, such as $\alpha$-shape \cite{alphashape} or composite metrics \cite{zhao2009}, could potentially improve prediction performance. Additionally, dimension reduction, for example by using principal component analysis (PCA) \cite{junttila2015}, would probably improve performance when using small training sets. Besides PCA, the deep belief network pretraining proposed in \cite{hinton2008} could be beneficial.

The prediction step of GPR is not computationally much more costly than using kNN. The most computationally expensive part is the GPR model training, which requires computing the matrix inverse of a large matrix (see the equations \eqref{predmean} and \eqref{predcov}). However, the matrix inverse can be precomputed for a given set of training data. After this, computing the prediction and the prediction interval only requires calculating matrix products. In the LOO case ($n_t=492$), computing the GPR prediction and intervals for a plot/cell took on average 345 ms in Matlab on a AMD Ryzen 1700X (3.4 GHz) processor, this is more expensive than kNN (1.5 ms), but still feasible for practice. For comparison, computing the Bayesian linear estimate took on average 18.5 s per plot. The training of GPR took 12.7 seconds.

The present work used fixed length scale, covariance function, and error magnitude, because finding the optimal values for these (hyper)parameters automatically is generally a nonconvex and computationally difficult optimization problem. The values, $l=10$ for the correlation length, and 10\% variance for the error $e$, were found by manual testing. The sub-optimal choice of these parameters might explain some of the tendency to produce too narrow prediction intervals for some stand attributes. Additionally, several commonly used covariance functions were tested;  Mat\'{e}rn $3/2$ covariance function was found to be the best performing. More advanced covariance functions, such as nonstationary covariance functions \cite{paciorek2004} or spectral mixture covariance functions \cite{wilson2013,wilson2016}, could potentially improve prediction accuracy. Further research is still needed on finding the optimal model formulation.

Due to extrapolation, GPR can produce unrealistic negative predictions. In this study, the negative predictions were corrected in a post-processing step. In the LOO cross-validation, total of 628 negative predictions occurred on 215 plots. Of these, 386 (61.5\%) occurred in cases where the corresponding field-measured value was zero (i.e. a missing tree species). Furthermore, 93\% of the negative predictions happened in cases where the corresponding field measurement was less than half of the average value of the stand attribute in the data set. The negative predictions thus occurred most commonly when predicting small stand attribute values. Additionally, 1039 cases where the field-measured stand attribute was zero were predicted to have a positive, non-zero value. Due to the more probable occurrence when predicting small values and the relatively symmetric distribution of the prediction error, the behavior of the negative predictions seems to be in line with the Gaussianity assumption in the GPR. 

In this study, GPR showed better reliability in all considered stand attribute predictions except the mean height of pine (the RMSE of basal area predictions of pine were practically the same for GPR and kNN) and the relative improvement of GPR predictions over the state-of-the-art method kNN were rather large being on average 4.6\%. This is contrary to the earlier studies where the reliability of ALS based forest inventory system has been examined by comparing different estimation methods. For example, Maltamo \emph{et al.} \cite{Maltamo2015} used visual pre-classification of aerial images to divide the study data into strata according to the main tree species and stand development stages. The aim was to improve species-specific estimates by applying more homogeneous reference data in kNN but the results were contradictory. The pre-classification did improve the accuracy of some species-specific stand attributes compared to the kNN estimates which applied whole study data as reference, but for some species-specific estimates the accuracy decreased. It is also notable that usually the accuracy of minor tree species did not improve, whereas in the present study the improvement was substantial especially for the minor species. 

Similar contradictory results have been obtained when comparing different statistical methods, such as neural networks or Bayesian approach \cite{niska2010neural,varvia}. For example, Niska \emph{et al.} \cite{niska2010neural} obtained more accurate species-specific volume estimates using neural networks at plot level than kNN but on the other hand kNN was more accurate on the stand level. R{\"a}ty \emph{et al.} \cite{raty2018} compared kNN estimates in which the species-specific estimates were obtained either by simultaneous imputation for all the species (as in this study) or by separate imputation for each species. The results concerning separate imputations were promising, but again, the results were contradictory.

\section{Conclusions}
In this article, the feasibility of Gaussian process regression for the estimation of species-specific stand attributes within the area based approach was evaluated. In addition to testing the prediction performance, the prediction credible intervals were also evaluated. GPR estimates were compared with a state-of-the-art kNN-based algorithm and a linear Bayesian inference based method. The effect of training set size on the performance was also examined.

The GPR estimates showed on average a 4.6\% relative improvement in RMSE over the reference kNN method in the leave-one-out cross-validation, generally smaller bias, and credible interval performance on par with the linear Bayesian inference. The GPR estimates kept the advantage even when tested using smaller training set sizes. Especially the credible interval performance proved robust with respect to the training set size.

The promising performance of GPR, the feasible computational cost, and that it provides prediction intervals make GPR an attractive method to use in forestry applications. Especially the plot level prediction uncertainty information provides many potential improvements in forest planning.

\bibliographystyle{IEEEtran}
\bibliography{IEEEabrv,bibliography}

\begin{IEEEbiography}[{\includegraphics[width=1in,height=1.25in,clip,keepaspectratio]{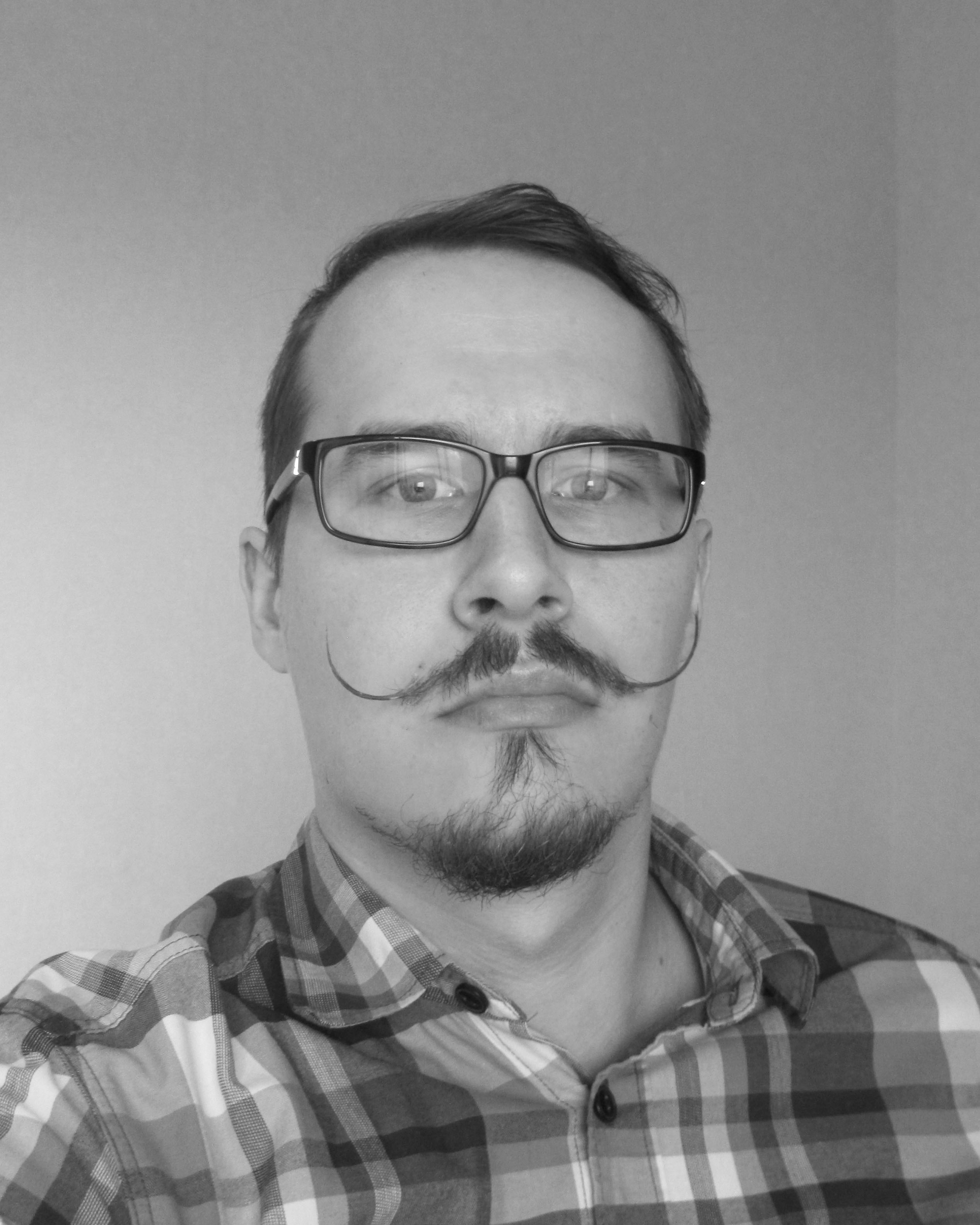}}]{Petri Varvia}
was born in Karttula, Finland, in 1988. He received M.Sc. and Ph.D. degrees in Applied Physics from the University of Eastern Finland in 2013 and 2018, respectively. He is currently a postdoctoral researcher at the Laboratory of Mathematics in the Tampere University of Technology. His scholarly interests include statistical inverse problems, Bayesian statistics and remote sensing. 
\end{IEEEbiography}

\begin{IEEEbiography}[{\includegraphics[width=1in,height=1.25in,clip,keepaspectratio]{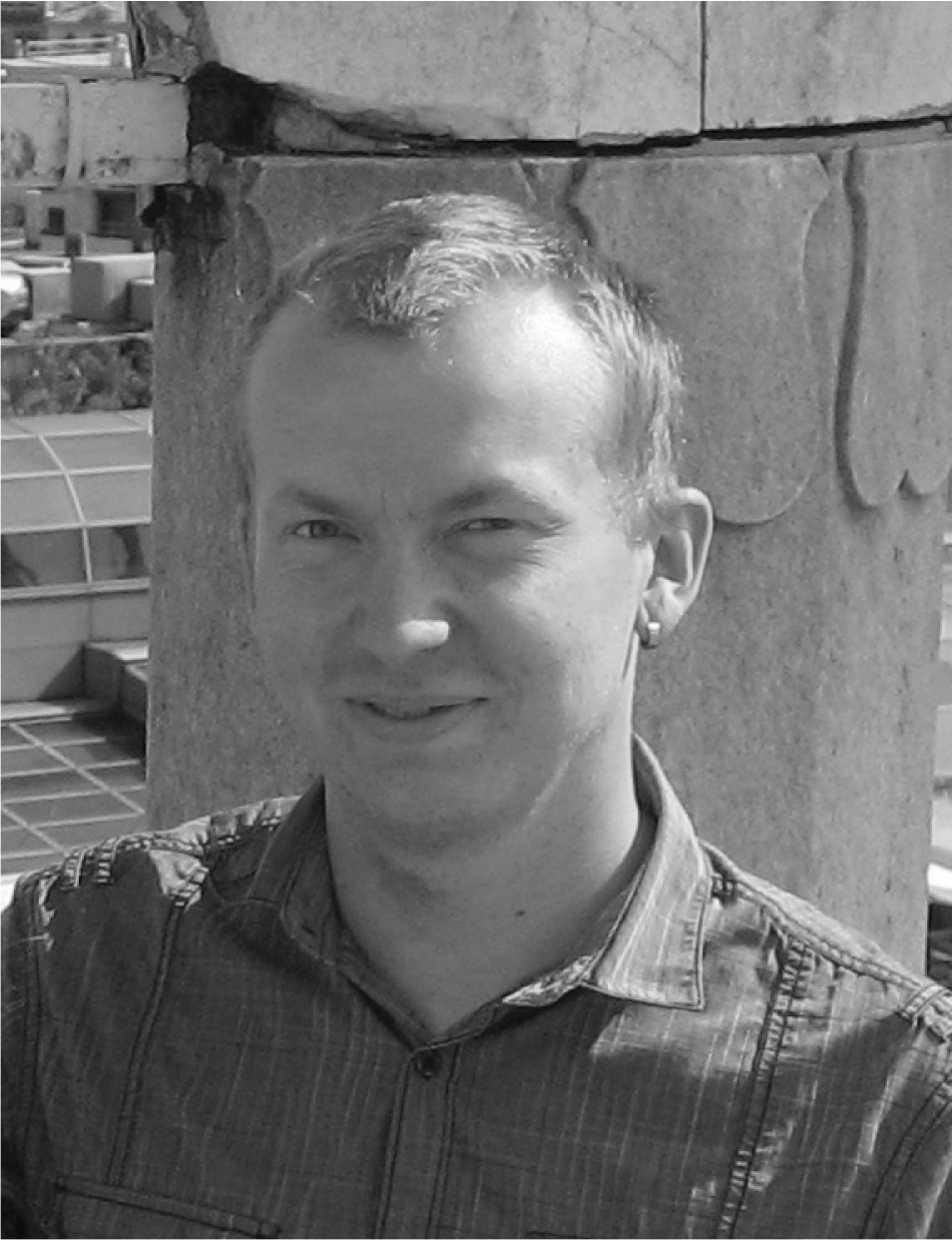}}]{Timo L\"ahivaara}
received the M.Sc. and Ph.D. degrees from the University of Kuopio, Finland, and University of Eastern Finland in 2006 and 2010, respectively.  Currently, he is a senior researcher at the Department of Applied Physics in the University of Eastern Finland.  His research interests are in computational wave problems and remote sensing.
\end{IEEEbiography}

\begin{IEEEbiography}[{\includegraphics[width=1in,height=1.25in,clip,keepaspectratio]{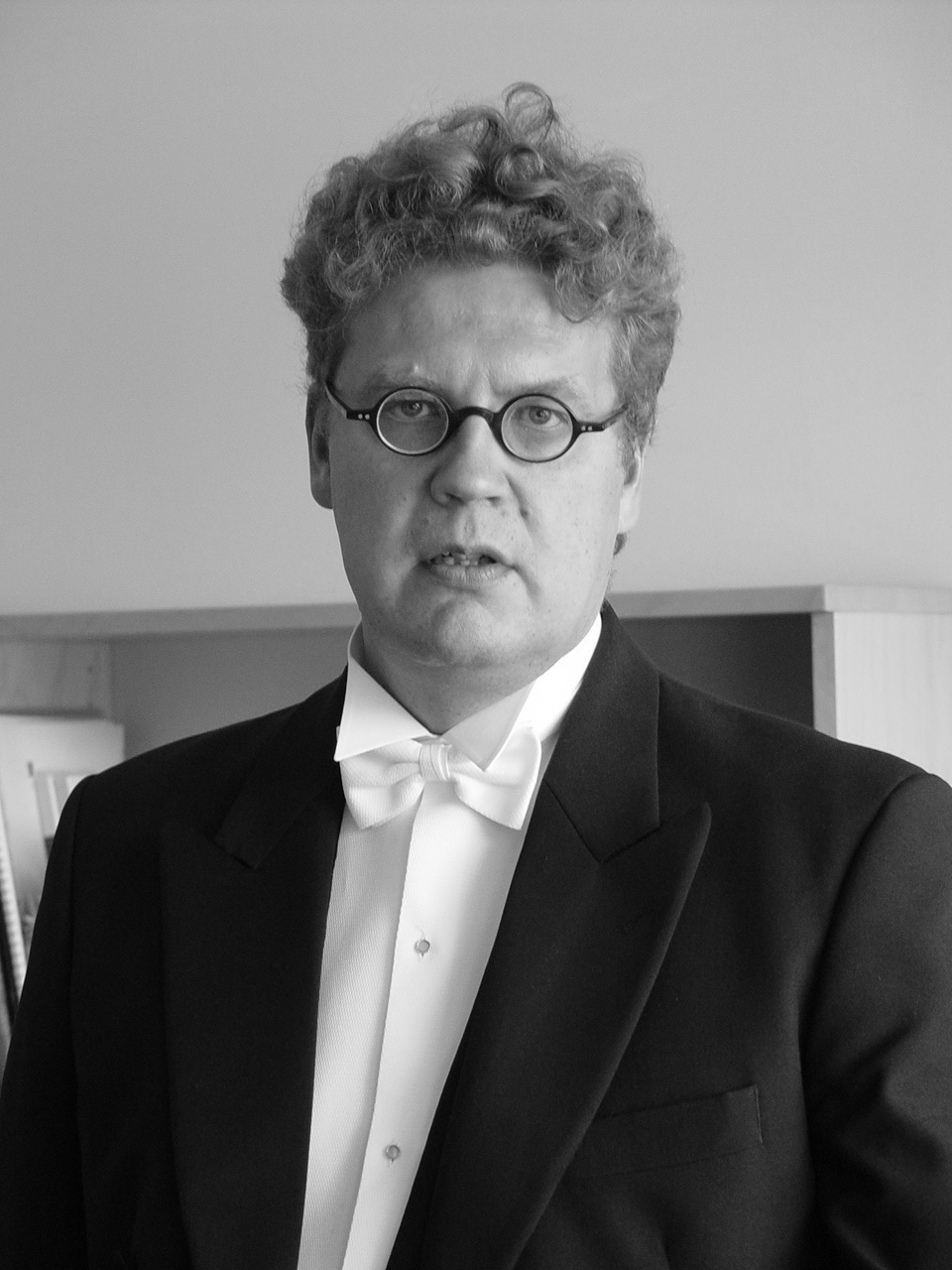}}]{Matti Maltamo}
was born in Jyv\"askyl\"a, Finland in 1965. He received the M.Sc., Lic.Sc., and D.Sc. degrees (with honors) in forestry from the University of Joensuu, Joensuu, Finland, in 1988, 1992, and 1998, respectively.

He is currently the Professor of Forest Mensuration Science with the Faculty of Science and Forestry, University of Eastern Finland. He has also worked as a visiting professor at the research group of professor Erik Naesset at the Norwegian University of Life Sciences. He was together with Naesset and Jari Vauhkonen the editor of the textbook “Forestry Applications of Airborne Laser Scanning -- concepts and case studies” published in 2014. He has published about 165 scientifically refereed papers. His specific research topic is Forestry Applications of ALS. He is an Associate Editor of the journal Canadian Journal of Forest Research.

Prof. Maltamo won together with professor Juha Hyypp\"a the First Innovation prize of the Finnish Society of Forest Science in 2010 about “Bringing airborne laser scanning to Finland. Maltamo also obtained bronze A.K. Cajander medal of the Finnish Society of Forest Science, 2012 
\end{IEEEbiography}

\begin{IEEEbiography}[{\includegraphics[width=1in,height=1.25in,clip,keepaspectratio]{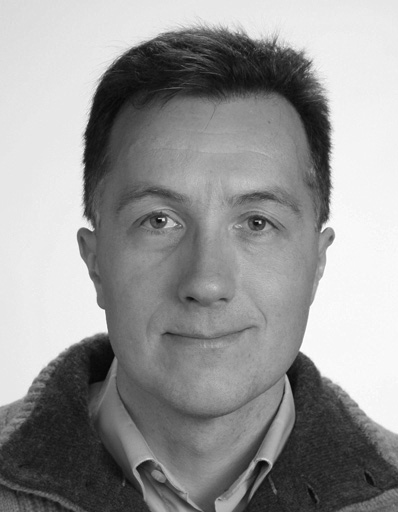}}]{Petteri Packalen}
was born in Rauma, Finland, in 1973. He received the M.Sc, Lic.Sc., and D.Sc. degrees in Forestry from the University of Joensuu, Joensuu, Finland, in 2002, 2007, and 2009, respectively. Currently, he is an Associate Professor in optimization of multi-functional forest management (Tenure Track) with the School of Forest Sciences, Faculty of Science and Forestry, University of Eastern Finland, Joensuu, Finland. Previously, he has been an Assistant, Senior Assistant, and Professor with the Faculty of Forestry, University of Joensuu. From August 2011 to July 2012, he was a Visiting Research Scientist at the Oregon State University, Corvallis, OR, USA. He has authored over 80 peer-reviewed research articles. Recently, his focus has been on time series, nearest neighbor imputation, combined use of ALS and spectral data in forest inventory, and the use of ALS in wildlife management. Since 2007, he has also been a Consultant for remote sensing-based forest inventory. His research interests include both practical and theoretical aspects of utilizing remote-sensing data in the monitoring and assessment of the forest environment. 
\end{IEEEbiography}

\begin{IEEEbiography}[{\includegraphics[width=1in,height=1.25in,clip,keepaspectratio]{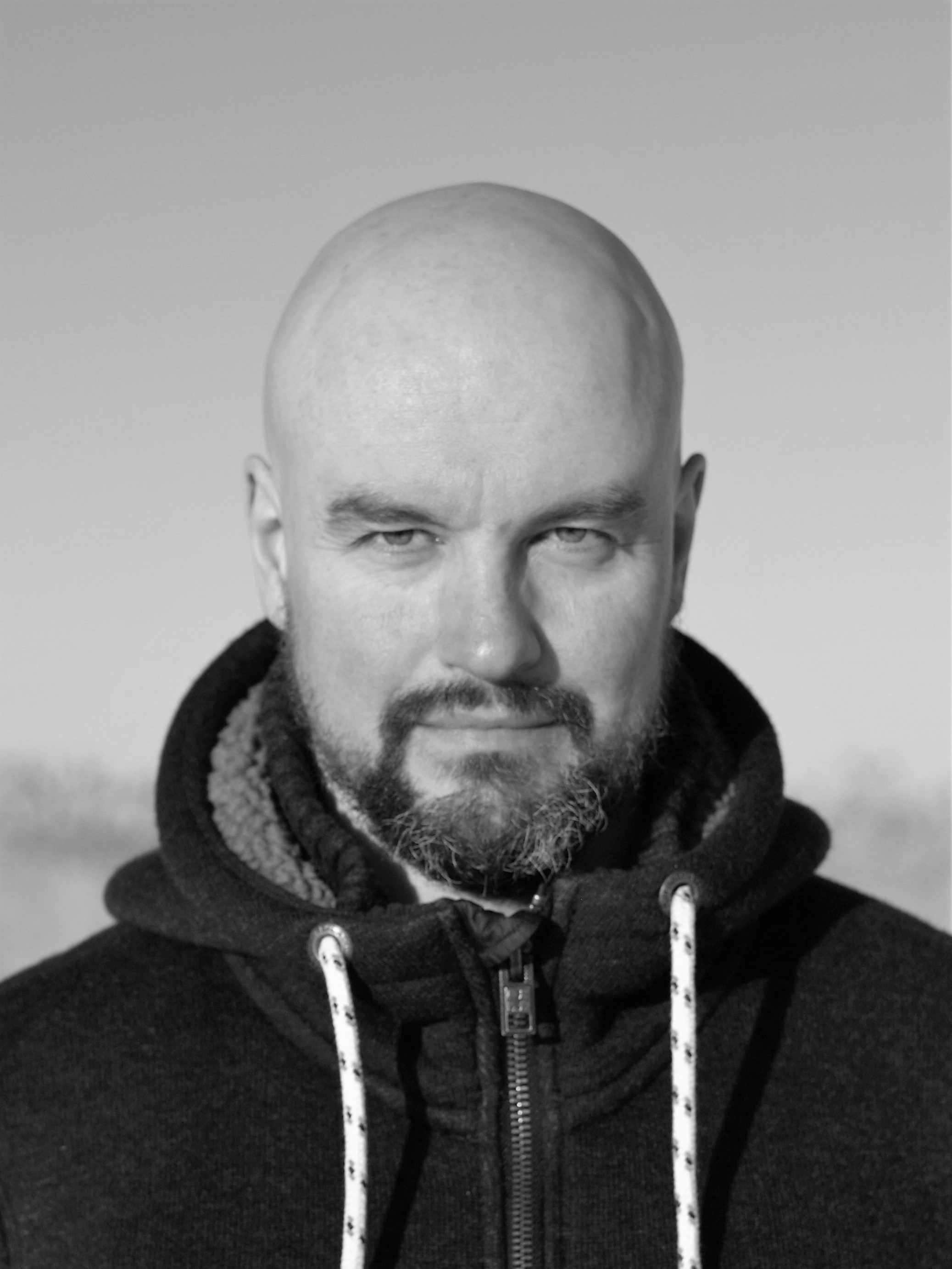}}]{Aku Sepp\"anen}
is an Associate Professor in the Department of Applied Physics at University of Eastern Finland, Kuopio. He received the M.Sc. and Ph.D. degrees from the University of Kuopio, Finland, in 2000 and 2006, respectively, and has authored 50 journal articles, 36 conference papers and 3 book chapters. His research interests are in statistical and computational inverse problems. He is one of the PIs in the Centre of Excellence in Inverse Modelling and Imaging (2018-2025) appointed by the Academy of Finland. The applications of his research include, e.g., industrial process imaging, non-destructive material testing and remote sensing of forest.
\end{IEEEbiography}

\end{document}